\documentclass[prd,onecolumn,nofootinbib]{revtex4}
\usepackage{bm,amsmath,amssymb,graphicx}

\begin{document}

\noindent
Classical and Quantum Gravity {\bf 41} (6), 065011 (2024)\\

\title{General-relativistic wave--particle duality with torsion}
\author{Francisco Ribeiro Benard Guedes$^{1,2}$}
\altaffiliation{frb2120@columbia.edu}
\author{Nikodem Janusz Pop{\l}awski$^1$}
\altaffiliation{NPoplawski@newhaven.edu}

\affiliation{$^{1}$Department of Mathematics and Physics, University of New Haven, West Haven, CT, USA}

\affiliation{$^{2}$Department of Biomedical Engineering, Columbia University, New York City, NY, USA}

\begin{abstract}
We propose that the four-velocity of a Dirac particle is related to its relativistic wave function by $u^i=\bar{\psi}\gamma^i\psi/\bar{\psi}\psi$.
This relativistic wave--particle duality relation is demonstrated for a free particle related to a plane wave in a flat spacetime.
For a curved spacetime with torsion, the momentum four-vector of a spinor is related to a generator of translation, given by a covariant derivative.
The spin angular momentum four-tensor of a spinor is related to a generator of rotation in the Lorentz group.
We use the covariant conservation laws for the spin and energy--momentum tensors for a spinor field in the presence of the Einstein--Cartan torsion to show that if the wave satisfies the curved Dirac equation, then the four-velocity, four-momentum, and spin satisfy the classical Mathisson--Papapetrou equations of motion.
We show that these equations reduce to the geodesic equation.
Consequently, the motion of a particle guided by the four-velocity in the pilot-wave quantum mechanics coincides with the geodesic motion determined by spacetime.
We also show how the duality and the operator form of the Mathisson--Papapetrou equations arise from the covariant Heisenberg equation of motion in the presence of torsion.
\end{abstract}
\maketitle

{\bf 1. Introduction}\\ \\
Wave--particle duality is the fundamental property of matter, which behaves under some conditions like waves and under other conditions like particles.
In quantum mechanics, a relativistic wave function propagates according to the Dirac equation, which explains the origin of the instrinsic angular momentum (spin) \cite{Dir}.
In classical mechanics, a relativistic particle with spin moves according to the Mathisson--Papapetrou equations of motion \cite{MP}.
Following wave--particle duality, the classical equations should correspond to the quantum equations.
Indeed, the motion of a classical pole-dipole test particle in general relativity can be derived from the equations of motion of the operators in the Heisenberg picture for a spinor wave function \cite{Wong}, the Foldy--Wouthuysen transformation applied to the Dirac equation \cite{Kan}, or the semiclassical expansion of the wave function as a power series in $\hbar$ \cite{Aud}.
The classical and quantum dynamics of Dirac particles in curved spacetime appear to be equivalent.
This equivalence has been demonstrated for a static gravitational field \cite{static}.

In this article, we propose a relativistic wave--particle duality relation between the four-velocity of a particle and the spinor wave function, analogous to that in the pilot-wave interpretation of quantum mechanics \cite{BH}.
We demonstrate the validity of this relation for a free particle related to a plane wave in flat spacetime (section 2). 
We use this relation to derive the Mathisson--Papapetrou equations from the covariant conservation laws for the spin and energy--momentum tensors for a spinor field, which satisfies the Dirac equation in a curved spacetime (section 3). 
For generality, we use the Einstein--Cartan theory of gravity \cite{EC,Niko}, in which the spacetime has both curvature and torsion.
We show that these equations reduce to the geodesic equation of motion, as in general relativity \cite{LL2}.
Finally, we show how the relativistic wave--particle duality relation and the operator form of the Mathisson--Papapetrou equations can be derived from the covariant Heisenberg equation of motion in the presence of torsion (section 4). 
These results provide a further validation of the equivalence between the classical and quantum dynamics of massive spin-1/2 particles in the gravitational field and a general-relativistic generalization of the pilot-wave theory (briefly discussed in section 5).\\ 

\noindent
{\bf Torsion}.\\
The affine connection $\Gamma^k_{ij}$ allows to construct the covariant derivative $\nabla_i$ of a vector that transforms under coordinate transformations like a tensor:
\[
\nabla_i V^k=\partial_i V^k+\Gamma^k_{ji}V^j,\quad \nabla_i V_k=\partial_i V_k-\Gamma^j_{ki}V_j.
\]
For tensors with more indices, each index produces a term with the affine connection.
The affine connection is not a tensor, but its variation is a tensor.
The antisymmetric part of the affine connection is also a tensor, referred to as the torsion tensor \cite{EC,Niko, Car}:
\begin{equation}
S^k_{\phantom{k}ij}=\frac{1}{2}(\Gamma^{k}_{ij}-\Gamma^{k}_{ji}).
\label{torsion}
\end{equation}

The metric tensor $g_{ik}$ allows to define a distance in curved coordinate system:
$ds^2=g_{ik}dx^i dx^k$, where $s$ is the interval.
The inverse metric tensor $g^{ik}$ satisfies $g_{ik}g^{jk}=\delta^j_i$, where $\delta^j_i$ is the Kronecker tensor.
Both metric tensors relate the contravariant and covariant vector indices: $V_i=g_{ik}V^k,\,V^i=g^{ik}V_k$.

The covariant derivative of the metric tensor is zero, which is called metricity: $\nabla_j g_{ik}=0$.
Therefore, the covariant differentiation commutes with converting between contravariant and covariant indices.
This relation determines the affine connection in terms of the metric tensor, its partial derivatives, and the torsion tensor \cite{EC,Niko}:
\[
\Gamma^{k}_{ij}=\mathring{\Gamma}^{k}_{ij}+S^k_{\phantom{k}ij}+S_{ij}^{\phantom{ij}k}+S_{ji}^{\phantom{ji}k},
\]
where the Christoffel symbols are given by
\[
\mathring{\Gamma}^{j}_{ik}=\frac{1}{2}g^{jl}(\partial_i g_{kl}+\partial_k g_{il}-\partial_l g_{ik}).
\]
They form a connection, referred to as the Levi-Civita connection.
The difference between the two connections is the contortion tensor $C^k_{\phantom{k}ij}=S^k_{\phantom{k}ij}+S_{ij}^{\phantom{ij}k}+S_{ji}^{\phantom{ji}k}$.

The curvature tensor is given by
\[
R^i_{\phantom{i}klm}=\partial_{l}\Gamma^{i}_{km}-\partial_{m}\Gamma^{i}_{kl}+\Gamma^{j}_{km}\Gamma^{i}_{jl}-\Gamma^{j}_{kl}\Gamma^{i}_{jm}.
\]
It can be decomposed into the Riemann tensor $\mathring{R}^i_{\phantom{i}klm}$, which is constructed from the $\mathring{\Gamma}^{j}_{ik}$ in the same way as the curvature tensor is constructed from $\Gamma^{j}_{ik}$, and a part that includes the contortion tensor: $R^i_{\phantom{i}klm}=\mathring{R}^i_{\phantom{i}klm}+\mathring{\nabla}_{l}C^i_{\phantom{i}km}-\mathring{\nabla}_{m}C^i_{\phantom{i}kl}+C^j_{\phantom{j}km}C^i_{\phantom{i}jl}-C^j_{\phantom{j}kl}C^i_{\phantom{i}jm}$, where $\mathring{\nabla}_{i}$ is the covariant derivative with respect to the Levi-Civita connection.\\

Torsion has a simple geometrical interpretation.
In the presence of curvature, a parallel transport (which also defines the covariant derivative) of a vector around a closed curve changes the vector.
In the presence of torsion, such a transport does not change the vector, but two parallel transports do not commute, which results from the following construction \cite{Schou}.
The parallel transport of an infinitesimal, four-dimensional vector $PR=dx^i$ from a point $P$ to an infinitesimally close point $Q$ such that $PQ=dx'^j$ adds to $dx^i$ a small correction: $\delta dx^i = -\Gamma^{i}_{jk}dx^j dx'^k$ \cite{LL2}.
After effecting the transport, the vector $dx^i+\delta dx^i$ points to a point $T$.
The parallel transport of the vector $dx'^i$ from a point $P$ to an infinitesimally close point $R$ adds to $dx^i$ a small correction: $\delta dx'^i = -\Gamma^{i}_{jk}dx^j dx'^k$.
After effecting the transport, the vector $dx'^i+\delta dx'^i$ points to a point $T'$.

\begin{figure}[th]
\centering
\includegraphics[width=1.2in]{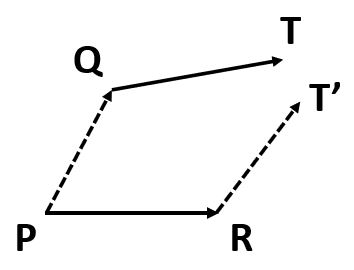}
\caption{Nonclosure of the parallelogram formed when two vectors are transported along each other depends on torsion.}
\label{parallel}
\end{figure}

Without torsion, points $T$ and $T'$ would coincide and form, together with points $P$, $Q$, and $R$, a parallelogram because $\delta dx'^i-\delta dx^i=\Gamma^{i}_{kj}dx^j dx'^k - \Gamma^{i}_{jk}dx^j dx'^k = 0$.
If the torsion tensor is not zero, however, the affine connection is asymmetric in the lower indices and $\delta dx'^i-\delta dx^i = -2S^i_{\phantom{i}jk}dx^j dx'^k$.
Points $T$ and $T'$ do not coincide, the distance between them is given by the torsion, the parallelogram is not closed, and the combination of two displacements of point $P$ (through $dx^i$ and $dx'^j$) depends on their order, as shown in Fig. \ref{parallel}.
Accordingly, the covariant derivatives of a scalar $\psi$ do not commute: $\nabla_i\nabla_j\psi-\nabla_j\nabla_i\psi=\partial_i\nabla_j\psi-\Gamma^{k}_{ji}\nabla_k\psi-\partial_j\nabla_i\psi+\Gamma^{k}_{ij}\nabla_k\psi=\partial_i\partial_j\psi-\partial_j\partial_i\psi+2S^k_{\phantom{k}ij}\nabla_k\psi=2S^k_{\phantom{k}ij}\nabla_k\psi$.
\\

\noindent
{\bf Dirac equation and spinors}.\\
The Dirac equation is a relativistic wave equation, which describes all massive spin-1/2 particles such as electrons and quarks \cite{Dir,QFT}.
This equation is consistent with the principles of quantum mechanics and special theory of relativity.
It predicts the existence of the spin angular momentum of particles and the existence of antiparticles, and accounts for the fine structure of the hydrogen spectrum.

In flat spacetime, the Dirac equation for a free particle with mass $m$ is given by
\begin{equation}
i\hbar\gamma^\mu\partial_\mu\psi=mc\psi,
\label{Diracflat}
\end{equation}
where $\psi$ is the four-component wave function and $\gamma^\mu$ are the $4\times4$ Dirac matrices, satisfying the relation
\[
\gamma^\mu\gamma^\nu+\gamma^\nu\gamma^\mu=2\eta^{\mu\nu}I_4,
\]
where $\eta^{\mu\nu}=\mbox{diag}(1,-1,-1,-1)$ is the Minkowski metric tensor and $I_4$ is the four-dimensional unit matrix.
Four is the lowest dimension for which the $\gamma^\mu$ can be defined.
In the presence of the electromagnetic field, $\partial_\mu$ is extended to $\partial_\mu+ieA_\mu$, where $A_\mu$ is the electromagnetic potential and $e$ is the electric charge of the particle.

In curved spacetime, the partial derivatives $\partial_\mu$ must be replaced with the covariant derivatives.
These derivatives can be determined in the spinor representation of the Lorentz group.
At every point in spacetime, in addition to a general coordinate system, it is possible to set up four linearly independent vectors $e^i_\mu$ such that $e^i_\mu e^k_\nu g_{ik}=\eta_{\mu\nu}$ \cite{EC,Niko}. 
This set of four vectors is referred to as a tetrad and allows to construct a locally flat geometry.
The tetrad relates the Greek vector indices of the locally flat, Lorentz coordinate system (of the special theory of relativity) to the Latin vector indices of the general coordinate system (of general relativity): $V^i=e^i_\mu V^\mu$.

The choice of the tetrad is not unique: a Lorentz transformation $\tilde{e}^i_\mu=\Lambda^\nu_\mu e^i_\nu$, where $\Lambda^\nu_\mu$ are the Lorentz matrices satisfying $\Lambda^\rho_\mu \Lambda^\sigma_\nu\eta_{\rho\sigma}=\eta_{\mu\nu}$, produces a new tetrad $\tilde{e}^i_\mu$.
Let $L$ be a 4$\times$4 matrix such that
\[
\gamma^\mu=\Lambda^\mu_\nu L\gamma^\nu L^{-1}.
\]
This condition represents the constancy of the Dirac matrices under a Lorentz tetrad rotation combined with a similarity transformation and gives the matrix $L$ as a function of the Lorentz matrix $\Lambda^\mu_\nu$.
For an infinitesimal Lorentz transformation $\Lambda^\mu_\nu=\delta^\mu_\nu+\epsilon^\mu_{\phantom{\mu}\nu}$, where $\epsilon_{\mu\nu}=-\epsilon_{\nu\mu}$ are infinitesimal quantities, the solution for $L$ is
\[
L=I_4+\frac{1}{2}\epsilon_{\mu\nu}G^{\mu\nu},\quad L^{-1}=I_4-\frac{1}{2}\epsilon_{\mu\nu}G^{\mu\nu},
\]
where $G^{\mu\nu}$ are the generators of the spinor representation of the Lorentz group \cite{Niko,QFT}:
\[
G^{\mu\nu}=\frac{1}{4}(\gamma^\mu\gamma^\nu-\gamma^\nu\gamma^\mu).
\]
The matrices $L$ compose the Lorentz group in spinor representation. 

A spinor $\psi$ and its adjoint $\bar{\psi}$ are defined as quantities that transform according to \cite{Niko,QFT}
\[
\tilde{\psi}=L\psi,\quad \tilde{\bar{\psi}}=\bar{\psi}L^{-1}.
\]
Accordingly, the product $\bar{\psi}\psi$ is a scalar: $\tilde{\bar{\psi}}\tilde{\psi}=\bar{\psi}\psi$.
The transformation law of the Dirac matrices shows that they can be regarded as quantities that have, in addition to the Lorentz vector index $\mu$, one spinor index and one adjoint-spinor index.
The product $\psi\bar{\psi}$ transforms like the Dirac matrices: $\tilde{\psi}\tilde{\bar{\psi}}=L\psi\bar{\psi}L^{-1}$.

The spinors $\psi$ and $\bar{\psi}$ can be used to construct bilinear forms that are linear both in $\psi$ and $\bar{\psi}$ and transform like tensors.
For example, $\bar{\psi}\gamma^\mu\psi$ transforms like a contravariant Lorentz vector: $\bar{\psi}\gamma^\mu\psi\rightarrow\bar{\psi}L^{-1}\Lambda^\mu_\nu L\gamma^\nu L^{-1}L\psi=\Lambda^\mu_\nu \bar{\psi}\gamma^\nu\psi$.

For an infinitesimal Lorentz transformation, the Hermitian conjugate of $L$ is $L^\dag=I_4+(1/8)\epsilon_{\mu\nu}(\gamma^{\nu\dag}\gamma^{\mu\dag}-\gamma^{\mu\dag}\gamma^{\nu\dag})$.
The relations $\gamma^{0\dag}=\gamma^0$ and $\gamma^{\alpha\dag}=-\gamma^\alpha$, where $\alpha\in\{1,2,3\}$, together with the definition of $\gamma^\mu$ give $L^\dag\gamma^0=\gamma^0 L^{-1}$.
Therefore, the quantity $\psi^\dag\gamma^0$ transforms like an adjoint spinor: $\psi^\dag\gamma^0\rightarrow\psi^\dag L^\dag\gamma^0=\psi^\dag\gamma^0 L^{-1}$.
Accordingly, we can associate the adjoint and the conjugate of a spinor:
\[
\bar{\psi}=\psi^\dag\gamma^0.
\]

\noindent
{\bf Covariant derivative of spinor}.\\
The spin connection is given by \cite{EC,Niko}
\[
\omega^\mu_{\phantom{\mu}\nu i}=e^\mu_k\omega^k_{\phantom{k}\nu i}=e^\mu_k \nabla_i e^k_\nu=e^\mu_k(\partial_i e^k_\nu+\Gamma^{k}_{ji}e^j_\nu).
\]
The spin connection transforms like a covariant vector under general coordinate transformations.
The spin connection allows to extend covariant differentiation to vectors with Lorentz indices:
\[
\nabla_i V^\mu=\partial_i V^\mu+\omega^\mu_{\phantom{\mu}\nu i}V^\nu,\quad \nabla_i V_\mu=\partial_i V_\mu-\omega^\nu_{\phantom{\nu}\mu i}V_\nu.
\]
For tensors with Lorentz indices, each index produces a term with the spin connection. 
Consequently, the covariant derivative of a tetrad is zero:
\[
\nabla_k e^i_\mu=\partial_k e^i_\mu+\Gamma^{i}_{jk}e^j_\mu-\omega^\nu_{\phantom{\nu}\mu k}e^i_\nu=0.
\]
Therefore, the covariant differentiation commutes with converting between coordinate and Lorentz indices.
This relation determines the spin connection in terms of the affine connection, the tetrad, and its partial derivatives.

The metricity of the affine connection leads to $\nabla_j g_{ik}=e^\mu_i e^\nu_k\nabla_j\eta_{\mu\nu}=-e^\mu_i e^\nu_k(\omega^\rho_{\phantom{\rho}\mu j}\eta_{\rho\nu}+\omega^\rho_{\phantom{\rho}\nu j}\eta_{\mu\rho})=-(\omega_{kij}+\omega_{ikj})=0$.
Consequently, $\nabla_i\eta_{\mu\nu}=0$ and the spin connection is antisymmetric in its first two indices: $\omega_{\mu\nu i}=-\omega_{\nu\mu i}$.

The derivative of a spinor does not transform like a spinor: $\partial_i\tilde{\psi}=L\partial_i\psi+\partial_i L\psi$.
Introducing the spinor connection $\Gamma_i$ that transforms according to
\[
\tilde{\Gamma}_i=L\Gamma_i L^{-1}+\partial_i LL^{-1}
\]
allows to construct the covariant derivative of a spinor:
\begin{equation}
\nabla_i\psi=\partial_i\psi-\Gamma_i \psi,
\label{covariant1}
\end{equation}
which transforms like a spinor: $\nabla_i\tilde{\psi}=\partial_i\tilde{\psi}-\tilde{\Gamma}_i\tilde{\psi}=L\partial_i\psi+\partial_i L\psi-(L\Gamma_i L^{-1}+\partial_i LL^{-1})L\psi=L\nabla_i\psi$ \cite{Niko}.
Because $\bar{\psi}\psi$ is a scalar, $\nabla_i(\bar{\psi}\psi)=\partial_i(\bar{\psi}\psi)$, the chain rule for covariant differentiation gives the covariant derivative of an adjoint spinor:
\begin{equation}
\nabla_i\bar{\psi}=\partial_i\bar{\psi}+\bar{\psi}\Gamma_i.
\label{covariant2}
\end{equation}

The Dirac matrices $\gamma^\mu$ transform like $\psi\bar{\psi}$, whose covariant derivative is $\nabla_i(\psi\bar{\psi})=
\nabla_i\psi\bar{\psi}+\psi\nabla_i\bar{\psi}=\partial_i(\psi\bar{\psi})-\Gamma_i \psi\bar{\psi}+\psi\bar{\psi}\Gamma_i=\partial_i(\psi\bar{\psi})-[\Gamma_i,\psi\bar{\psi}]$.
Therefore, the covariant derivative of a Dirac matrix is \cite{Niko}
\[
\nabla_i\gamma^\mu=\omega^{\mu}_{\phantom{\mu}\nu i}\gamma^\nu-[\Gamma_i,\gamma^\mu].
\]
A quantity $\bar{\psi}\gamma^i\nabla_i\psi$, where $\gamma^i=e^i_\mu\gamma^\mu$, transforms under Lorentz rotations (and general coordinate transformations) like a scalar:
\[
\bar{\psi}\gamma^i\nabla_i\psi\rightarrow\bar{\psi}L^{-1}L\gamma^i L^{-1}L\nabla_i\psi=\bar{\psi}\gamma^i\nabla_i\psi.
\]
So does a quantity $\nabla_i\bar{\psi}\gamma^i\psi$.
One can show that the difference $\bar{\psi}\gamma^i\nabla_i\psi-\nabla_i\bar{\psi}\gamma^i\psi$ is imaginary.
It can be used to constructed the simplest Lagrangian density for a Dirac field that contains its covariant derivatives.

The relation $\nabla_i\eta_{\mu\nu}=0$ gives $\nabla_i\gamma^\mu=0$ because the Dirac matrices $\gamma^\mu$ depend only on the tensor $\eta_{\mu\nu}$.
Consequently, $\gamma_\mu\nabla_i\gamma^\mu=\omega_{\mu\nu i}\gamma^\mu \gamma^\nu-\gamma_\mu \Gamma_i \gamma^\mu+4\Gamma_i=0$.
Its solution is $\Gamma_i=-\frac{1}{4}\omega_{\mu\nu i}\gamma^\mu \gamma^\nu-A_i I_4$, where $A_i$ is a covariant vector, which could be proportional to the electromagnetic potential.
Therefore, the spinor connection $\Gamma_i$ is given, up to the addition of an arbitrary vector multiple of the unit matrix, by the Fock--Ivanenko coefficients \cite{Niko,FI}:
\[
\Gamma_i=-\frac{1}{4}\omega_{\mu\nu i}\gamma^\mu \gamma^\nu=-\frac{1}{2}\omega_{\mu\nu i}G^{\mu\nu}.
\]

The curvature tensor with two Lorentz and two coordinate indices depends on the spin connection and its partial derivatives:
\[
R^\mu_{\phantom{\mu}\nu ik}=\omega^\mu_{\phantom{\mu}\nu k,i}-\omega^\mu_{\phantom{\mu}\nu i,k}+\omega^\rho_{\phantom{\rho}\nu k}\omega^\mu_{\phantom{\mu}\rho i}-\omega^\rho_{\phantom{\rho}\nu i}\omega^\mu_{\phantom{\mu}\rho k}.
\]
The commutator of the covariant derivatives of a spinor is
\[
\nabla_i\nabla_j\psi-\nabla_j\nabla_i\psi=K_{ij}\psi+2S^k_{\phantom{k}ij}\nabla_k\psi,
\]
where
\[
K_{ij}=\partial_j\Gamma_i-\partial_i\Gamma_j+[\Gamma_i,\Gamma_j]=\frac{1}{4}R_{\mu\nu ij}\gamma^\mu \gamma^\nu=\frac{1}{2}R_{\mu\nu ij}G^{\mu\nu}
\]
is the curvature spinor, which transforms like a quantity with one spinor index and one adjoint-spinor index: $\tilde{K}_{ij}=LK_{ij}L^{-1}$ \cite{Niko}.\\

\noindent
{\bf Dirac equation in curved spacetime}.\\
The Dirac Lagrangian density for a spinor field, representing a particle in a gravitational field, is given by
\begin{equation}
\mathfrak{L}_\psi=\frac{1}{2}i\hbar\mathfrak{e}e^i_\mu(\bar{\psi}\gamma^\mu\nabla_i\psi-\nabla_i\bar{\psi}\gamma^\mu\psi)-mc\mathfrak{e}\bar{\psi}\psi,
\label{Lagrangian}
\end{equation}
where $\mathfrak{e}$ is the determinant of the tetrad $e^i_\mu$.
Varying the quantity (\ref{Lagrangian}) with respect to $\bar{\psi}$ and $\psi$, respectively, gives the Dirac equation and its adjoint \cite{Niko,FI}:
\begin{equation}
i\hbar \gamma^\mu e_\mu^i \nabla_i \psi = mc\psi,\quad -i\hbar \nabla_i\bar{\psi} \gamma^\mu e_\mu^i= mc\bar{\psi}.
\label{Dirac}
\end{equation}
These equations generalize the Dirac equation (\ref{Diracflat}) to a curved spacetime, in which the covariant derivatives (\ref{covariant1}) and (\ref{covariant2}) are used.
In terms of the Fock--Ivanenko coefficients, this equation is
\[
i\hbar \gamma^\mu e_\mu^i (\partial_i \psi + \frac{1}{4}\omega_{\rho\sigma i}\gamma^\rho \gamma^\sigma) = mc\psi.
\]

Subtracting the first equation in (\ref{Dirac}) multiplied by $\bar{\psi}$ from the second equation multiplied by $\psi$ gives the conservation law for the Dirac current density $\mathfrak{j}^k=\mathfrak{e}\bar{\psi}\gamma^k\psi$:
\begin{equation}
\bar{\psi}\gamma^k\nabla_k\psi+\nabla_k\bar{\psi}\gamma^k\psi=\nabla_k(\bar{\psi}\gamma^k\psi)=\frac{1}{\mathfrak{e}}\nabla_k(\mathfrak{e}\bar{\psi}\gamma^k\psi)=0.
\label{current}
\end{equation}

\noindent
{\bf Energy--momentum and spin densities}.\\
Varying the Lagrangian density (\ref{Lagrangian}) with respect to the spin connection gives the spin density $\mathfrak{S}_{\mu\nu}^{\phantom{\mu\nu}i}=-\mathfrak{S}_{\nu\mu}^{\phantom{\nu\mu}i}=2\delta\mathfrak{L}_\psi/\delta\omega^{\mu\nu}_{\phantom{\mu\nu}i}$ for a spinor field \cite{Niko}:
\begin{equation}
\mathfrak{S}_{\mu\nu}^{\phantom{\mu\nu}i}=\frac{1}{8}i\hbar\mathfrak{e}\bar{\psi}\{\gamma^i,[\gamma_\mu,\gamma_\nu]\}\psi=\frac{1}{2}i\hbar\mathfrak{e}\bar{\psi}\{\gamma^i,G_{\mu\nu}\}\psi.
\label{spindensity}
\end{equation}
This quantity does not depend on the spinor mass.
The spin tensor is equal to the spin density divided by $\mathfrak{e}$.
Varying the Lagrangian density (\ref{Lagrangian}) with respect to the tetrad gives the tetrad (canonical) energy--momentum density $\mathfrak{T}^{\phantom{i}\mu}_i=\delta\mathfrak{L}_\psi/\delta e^i_\mu$ for a spinor field:
\begin{equation}
\mathfrak{T}^{\phantom{i}\mu}_i=\frac{1}{2}i\hbar\mathfrak{e}(\bar{\psi}\gamma^\mu\nabla_i\psi-\nabla_i\bar{\psi}\gamma^\mu\psi-e^\mu_i\bar{\psi}\gamma^j\nabla_j\psi+e^\mu_i\nabla_j\bar{\psi}\gamma^j\psi)+mc\mathfrak{e}e^\mu_i\bar{\psi}\psi.
\label{energymomentumdensity}
\end{equation}
The energy--momentum tensor is equal to the energy--momentum density divided by $\mathfrak{e}$.

The conservation law for spin density is \cite{Niko}
\begin{equation}
\nabla_k\mathfrak{S}_{ij}^{\phantom{ij}k}-2S_k\mathfrak{S}_{ij}^{\phantom{ij}k}=\mathfrak{T}_{ij}-\mathfrak{T}_{ji},
\label{spinconservation}
\end{equation}
where $S_k=S^i_{\phantom{i}ki}$.
Without spin, the energy--momentum tensor is symmetric, as in general relativity.
The conservation law for tetrad energy--momentum density is
\begin{equation}
\nabla_j\mathfrak{T}^{\phantom{i}j}_{i}-2S_j \mathfrak{T}^{\phantom{i}j}_i=2S^j_{\phantom{j}ki}\mathfrak{T}^{\phantom{j}k}_j+\frac{1}{2}R^{kl}_{\phantom{kl}ji}\mathfrak{S}_{kl}^{\phantom{kl}j}.
\label{energymomentumconservation}
\end{equation}
The conservation law (\ref{energymomentumconservation}) applied to the energy--momentum density (\ref{energymomentumdensity}) gives the Dirac equations (\ref{Dirac}).\\

\noindent
{\bf Einstein--Cartan theory}.\\
The Einstein--Cartan theory, formulated by Sciama and Kibble \cite{EC}, is the simplest theory of gravity that extends general relativity by relaxing the symmetry condition of the affine connection.
In this theory, the Lagrangian density for the gravitational field is given by $\mathfrak{L}_g=(-1/2\kappa)\mathfrak{e}R$, where $R=R_{ik}g^{ik}$ is the Ricci scalar, $R_{ik}=R^j_{\phantom{j}ijk}$ is the Ricci tensor, and $\kappa=8\pi G/c^4$.
In general relativity, the curvature tensor $R^l_{\phantom{l}ijk}$ reduces to the Riemann tensor $\mathring{R}^l_{\phantom{l}ijk}$.

The field equations of the Einstein--Cartan theory are obtained from the principle of least action with the variations of the metric and torsion tensors.
Equivalently, the field equations can be derived from the variations of the tetrad and spin connection.
Equaling the variation of the total Lagrangian density for the gravitational field and matter with respect to the spin connection to zero gives the Cartan equations \cite{EC,Niko}:
\begin{equation}
S^i_{\phantom{i}\mu\nu}-S_\mu e^i_\nu+S_\nu e^i_\mu=-\frac{\kappa}{2\mathfrak{e}}\mathfrak{S}^{\phantom{\mu\nu}i}_{\mu\nu}.
\label{Cartan}
\end{equation}
Equaling the variation of the total Lagrangian density for the gravitational field and matter with respect to the tetrad to zero gives the Einstein equations:
\[\
R^\mu_{\phantom{\mu}i}-\frac{1}{2}Re^\mu_i=\frac{\kappa}{\mathfrak{e}}\mathfrak{T}^{\phantom{i}\mu}_i.
\]
The Bianchi identity $\nabla_{[l}R^i_{\phantom{i}|n|jk]}=2R^i_{\phantom{i}nm[j}S^m_{\phantom{m}kl]}$ and the cyclic identity $R^m_{\phantom{m}[jkl]}=-2\nabla_{[l}S^m_{\phantom{m}jk]}+4S^m_{\phantom{m}n[j}S^n_{\phantom{n}kl]}$, where the square brackets denote antisymmetrization of indices except those between the bars, together with the Einstein
and Cartan field equations give the conservation laws for the spin density (\ref{spinconservation}) and for the tetrad energy--momentum density (\ref{energymomentumconservation}) \cite{Niko}.

These field equations relate the curvature of spacetime to the energy and momentum (\ref{energymomentumdensity}) of matter and the torsion of spacetime to the spin angular momentum (\ref{spindensity}) of matter.
According to the Cartan equations, the torsion tensor is proportional to the spin density.
In the absence of spin, the torsion tensor is therefore zero and the affine connection reduces to the Levi-Civita connection, given by the Christoffel symbols.
Consequently, the Einstein--Cartan theory reduces to the general theory of relativity.
This theory is also indistinguishable in predictions from general relativity at densities of matter that are lower than the Cartan density ($\sim 10^{45}$kg/m$^3$), so it passes all observational and experimental tests of general relativity \cite{EC}.\footnote{
Extending general relativity to the Einstein--Cartan theory may solve several problems in quantum field theory and cosmology.
Torsion imposes a spatial extension on fermions \cite{non} and removes the ultraviolet divergence of radiative corrections represented by loop Feynman diagrams \cite{torsional}.
Torsion also generates a negative correction from the spin-torsion coupling to the energy density, which acts like gravitational repulsion, which may prevent the formation of singularities in black holes and at the beginning of the Universe \cite{reg}.
Consequently, the collapsing matter in a black hole would avoid a singularity and instead reach a nonsingular bounce, after which it would expand as a new, closed universe on the other side of its event horizon \cite{universe}.
Accordingly, our Universe might have originated as a baby universe in a parent black hole existing in another universe.}\\

{\bf 2. Four-velocity of spinor}\\ \\
\noindent
{\bf Velocity of spinor}.\\
We aim to demonstrate that a relativistic spinor wave, described by the Dirac equation, can be associated with a particle of the same mass according to
\begin{equation}
u^i=\frac{\bar{\psi}\gamma^i\psi}{\bar{\psi}\psi},
\label{velocity}
\end{equation}
where $u^i=dx^i/ds$ is the four-velocity of the particle and $s$ is the affine parameter along a world line of the particle.
This relation describes the relativistic wave--particle duality.
It is equal to the ratio of the Dirac density four-current and the Dirac density scalar. 
In the pilot-wave interpretation of quantum mechanics, this relation coincides with the Bohm--Hiley four-velocity of a particle guided by the wave function \cite{BH}.

It is straightforward to demonstrate that the relation (\ref{velocity}) is satisfied for a free particle associated with a plane spinor wave in a flat spacetime (in the absence of torsion) \cite{duality}.
The Dirac equation (\ref{Diracflat}) can be written in the Hamiltonian form:
\[
i\hbar\frac{\partial\psi}{\partial t}=-i\hbar{\bm\alpha}\cdot{\bm\nabla}\psi+mc\beta\psi,
\]
where ${\bm\alpha}$ is the vector formed from the matrices $\alpha^\mu=\beta\gamma^\mu$ with the space indices $\mu$, and $\beta=\gamma^0$.
If the particle has four-momentum $p_\mu$, the corresponding spinor wave function has a form of a plane wave proportional to $\exp(-ip_\mu x^\mu/\hbar)$, where $x^\mu$ are the spacetime coordinates.
Consequently, the Dirac equation becomes
\[
E\psi={\bf p}\cdot{\bm\alpha}c\psi+mc^2\beta\psi,
\]
where ${\bf p}=m{\bf v}\gamma$ is the momentum of the particle and $E=mc^2\gamma$ is its energy, satisfying $E^2=({\bf p}c)^2+(mc^2)^2$, with velocity ${\bf v}$ and $\gamma=(1-{\bf v}^2/c^2)^{-1/2}$.
In the Dirac representation, $\gamma^0=\left( \begin{array}{cc}
    I_2 & 0 \\
    0 & -I_2 \end{array} \right), \gamma^\mu=\left( \begin{array}{cc}
    0 & \sigma^\mu \\
    -\sigma^\mu & 0 \end{array} \right), \alpha^\mu=\left( \begin{array}{cc}
    0 & \sigma^\mu \\
    \sigma^\mu & 0 \end{array} \right)$, where $I_2$ is the two-dimensional unit matrix, $\sigma^\mu$ are the Pauli matrices, and $\mu=1,2,3$.
The normalized plane-wave solution of the Dirac equation is
\begin{equation}
    \psi({\bf r},t)=\left( \begin{array}{cc}
    (E+mc^2)I_2 & {\bm\sigma}\cdot{\bf p}c \\
    {\bm\sigma}\cdot{\bf p}c & (E+mc^2)I_2 \end{array} \right)\left( \begin{array}{c}
    \xi \\
    \eta \end{array} \right)\frac{1}{\sqrt{2mc^2(E+mc^2)}}\exp[i({\bf p}\cdot{\bf r}-Et)/\hbar],
    \label{plane}
\end{equation}
where ${\bm\sigma}$ is the vector formed from the Pauli matrices, the square matrix is the spinor representation of the boost from rest to the velocity ${\bf v}={\bf p}c^2/E$ \cite{Niko,QFT}, and $\xi$ and $\eta$ are two-dimensional (up and down) spinors describing positive (particle) and negative (antiparticle) energy states.

For a spin-up, positive-energy state, $\xi=\left( \begin{array}{c}
    1 \\
    0 \\ \end{array} \right)$ and $\eta=\left( \begin{array}{c}
    0 \\
    0 \\ \end{array} \right)$.
The scalar bilinear composed from the plane wave (\ref{plane}) is
\begin{eqnarray}
    & & \bar{\psi}\psi=\psi^\dagger\gamma^0\psi=\frac{1}{2mc^2(E+mc^2)}\Bigl[(E+mc^2)^2-(1,0)({\bm\sigma}\cdot{\bf p})({\bm\sigma}\cdot{\bf p})c^2
    \left( \begin{array}{c}
    1 \\
    0 \\ \end{array} \right)\Bigr] \nonumber \\
    & & =\frac{(E+mc^2)^2-{\bf p}^2 c^2}{2mc^2(E+mc^2)}=1,
    \nonumber
\end{eqnarray}
as expected.
The vector bilinear components composed from (\ref{plane}), using the relation $p^\mu=mcu^\mu$ for a free particle and an identity $\sigma^\mu\sigma^\nu+\sigma^\nu\sigma^\mu=-2\eta^{\mu\nu}I_2$, are
\begin{eqnarray}
    & & \bar{\psi}\gamma^0\psi=\psi^\dagger\psi=\frac{1}{2mc^2(E+mc^2)}\Bigl[(E+mc^2)^2+(1,0)({\bm\sigma}\cdot{\bf p})({\bm\sigma}\cdot{\bf p})c^2
    \left( \begin{array}{c}
    1 \\
    0 \\ \end{array} \right)\Bigr] \nonumber \\
    & & =\frac{(E+mc^2)^2+{\bf p}^2 c^2}{2mc^2(E+mc^2)}=\frac{E}{mc^2}=u^0,
    \nonumber
\end{eqnarray}
and for the space components:
\begin{eqnarray}
    & & \bar{\psi}\gamma^\mu\psi=\psi^\dagger\alpha^\mu\psi=\frac{E+mc^2}{2mc^2(E+mc^2)}\Bigl[(1,0)\sigma^\mu({\bm\sigma}\cdot{\bf p})c\left( \begin{array}{c}
    1 \\
    0 \\ \end{array} \right)+(1,0)({\bm\sigma}\cdot{\bf p})c\sigma^\mu\left( \begin{array}{c}
    1 \\
    0 \\ \end{array} \right)\Bigr] \nonumber \\
    & & =\frac{p^\mu}{mc}=u^\mu.
    \nonumber
\end{eqnarray}
These bilinears validate the relation (\ref{velocity}) for a free spinor particle in a flat spacetime.
Similar calculations can be carried out for a spin-down or a negative-energy state.

The normalization of the four-velocity (\ref{velocity}), $u^\mu u_\mu=1$, follows from $p^\mu p_\mu=m^2 c^2$.
Consequently, this four-velocity is a timelike four-vector.
This normalization also holds in curved spacetime because at the location of the particle one can construct a locally flat system of coordinates \cite{LL2} and because the normalization is a scalar relation independent of the choice of the coordinates.
The four-velocity defined by (\ref{velocity}) therefore  remains well-behaved and is therefore a meaningful physical quantity that can be used to define de Broglie--Bohm trajectories for a spin-1/2 particle.

For the Klein--Gordon equation describing a spin-0 particle \cite{QFT}, the four-velocity defined as the ratio of the density four-current and the density scalar analogously to (\ref{velocity}) could be spacelike (superluminal) and thus without particle interpretation.
However, a representation analogous to the spin-1/2 Foldy--Wouthuysen transformation \cite{QFT,FW}, in which the spin-0 particle and antiparticle are decoupled in time dynamics, admits a positive particle density and the corresponding density current, for which the four-velocity is subluminal and thus well-behaved, allowing to define trajectories \cite{super}.
The Foldy--Wouthuysen representation is not needed in this work; the relation (\ref{velocity}) is already satisfied in the Dirac representation.\\

\noindent
{\bf Momentum and spin of spinor}.\\
The conservation of the momentum four-vector $P_i$ follows from the symmetry of a system under spacetime translations.
The four-momentum operator $\hat{P}_i$ of a spinor is therefore associated with a generator of translation, which in curved spacetime is given by a covariant derivative \cite{Wong}:
\begin{equation}
\hat{P}_i\psi=i\hbar\nabla_i\psi,\quad \hat{P}_i\bar{\psi}=-i\hbar\nabla_i\bar{\psi}.
\label{momentum}
\end{equation}
The conservation of the intrinsic angular momentum (spin) four-tensor $S_{ik}$ follows from the symmetry of a system under spacetime rotations.
The spin four-tensor operator $\hat{S}_{ik}$ of a spinor is therefore associated with a generator of rotation $G_{ik}$:
\begin{equation}
\hat{S}_{ik}\psi=i\hbar G_{ik}\psi,\quad \hat{S}_{ik}\bar{\psi}=i\hbar \bar{\psi}G_{ik},
\label{spin}
\end{equation}
following $\hat{S}_{ik}\bar{\psi}=\hat{S}_{ik}\psi^\dagger\gamma^0=(iG_{ik}\psi)^\dagger\gamma^0=-i\psi^\dagger G_{ik}^\dagger\gamma^0=i\psi^\dagger\gamma^0 G_{ik}=i\bar{\psi}G_{ik}$.

We consider a free (without fields other than gravity) particle with definite values of the four-momentum and spin, which are respectively eigenvalues of their operators:
\[
\hat{P}_i\psi=P_i\psi,\quad \hat{S}_{ik}\psi=S_{ik}\psi.
\]
In this case, the relations (\ref{momentum}) and (\ref{spin}) can be used with the operators replaced with the respective eigenvalues.
The mass $m$ of the particle is given by \cite{Niko}
\begin{equation}
mc=u^i P_i.
\label{mass}
\end{equation}
Using (\ref{momentum}), the Dirac equations (\ref{Dirac}) can be written as
\[
P_i \gamma^i\psi=mc\psi,\quad \bar{\psi}\gamma^i P_i=mc\bar{\psi},
\]
giving
\[
P_i\frac{\bar{\psi}\gamma^i \psi}{\bar{\psi}\psi}=mc.
\]
Comparing this relation with (\ref{mass}) validates the formula (\ref{velocity}) for the four-velocity of a spinor.

Substituting the eigenvalue form of (\ref{momentum}) into the tetrad energy--momentum density (\ref{energymomentumdensity}) gives
\[
\mathfrak{T}^{\phantom{i}\mu}_i=\mathfrak{e}(P_i\bar{\psi}\gamma^\mu\psi-e^\mu_i P_j\bar{\psi}\gamma^j\psi)+mc\mathfrak{e}e^\mu_i\bar{\psi}\psi.
\]
Using (\ref{velocity}), this density becomes
\begin{eqnarray}
& & \mathfrak{T}^{\phantom{i}\mu}_i=\mathfrak{e}\bar{\psi}\psi(P_i u^\mu-e^\mu_i P_j u^j)+mc\mathfrak{e}e^\mu_i\bar{\psi}\psi=\mathfrak{e}\bar{\psi}\psi P_i u^\mu, 
\nonumber \\
& & \mathfrak{T}^{\phantom{i}k}_i=\mathfrak{e}\bar{\psi}\psi P_i u^k.
\label{franikoenergymomentum}
\end{eqnarray}
Substituting the eigenvalue form of (\ref{spin}) into the spin density (\ref{spindensity}) gives
\[\
\mathfrak{S}_{\mu\nu}^{\phantom{\mu\nu}i}=\frac{1}{2}i\hbar\mathfrak{e}\bar{\psi}(\gamma^i G_{\mu\nu}+G_{\mu\nu}\gamma^i)\psi=\mathfrak{e}\bar{\psi}\gamma^i\psi S_{\mu\nu}.
\]
Using (\ref{velocity}), this density becomes
\begin{equation}
\mathfrak{S}_{jk}^{\phantom{jk}i}=\mathfrak{e}\bar{\psi}\psi S_{jk}u^i.
\label{franikospin}
\end{equation}

The generators satisfy $\gamma^\nu G_{\mu\nu}=0$.
Contracting the spin four-tensor in (\ref{spin}) with the four-velocity (\ref{velocity}) therefore gives
\begin{equation}
S_{ik}u^k=S_{ik}\frac{\bar{\psi}\gamma^k \psi}{\bar{\psi}\psi}=\frac{\bar{\psi}\gamma^k S_{ik}\psi}{\bar{\psi}\psi}=i\hbar\frac{\bar{\psi}\gamma^k G_{ik}\psi}{\bar{\psi}\psi}=0.
\label{orthogonality}
\end{equation}
The spin four-tensor is orthogonal to the four-velocity.\\

{\bf 3. Equations of motion for spinor}\\ \\
\noindent
{\bf Conservation laws and equations of motion}.\\
The relations (\ref{franikospin}) and (\ref{orthogonality}) give $\mathfrak{S}_{jk}^{\phantom{jk}k}=0$.
Consequently, the Cartan equations (\ref{Cartan}) give
\begin{equation}
S_i=0.
\label{torsionvector}
\end{equation}
Putting (\ref{franikoenergymomentum}) and (\ref{franikospin}) into the conservation law (\ref{spinconservation}) for spin density with (\ref{torsionvector}) gives
\[
\nabla_k(\mathfrak{e}\bar{\psi}\psi S_{ij}u^k)=\mathfrak{e}\bar{\psi}\psi(P_i u_j-P_j u_i).
\]
Using $u^k\nabla_k=D/ds$, where $D$ denotes the covariant differential, the four-velocity (\ref{velocity}), and the conservation of current (\ref{current}), this relation can be written as
\begin{eqnarray}
& & \mathfrak{e}\bar{\psi}\psi(P_i u_j-P_j u_i)=\mathfrak{e}\bar{\psi}\psi u^k\nabla_k S_{ij}+S_{ij}\nabla_k(\mathfrak{e}\bar{\psi}\psi u^k) \nonumber \\
& & =\mathfrak{e}\bar{\psi}\psi\frac{DS_{ij}}{ds}+S_{ij}\nabla_k(\mathfrak{e}\bar{\psi}\gamma^k\psi)=\mathfrak{e}\bar{\psi}\psi\frac{DS_{ij}}{ds}, \nonumber
\end{eqnarray}
which is equivalent to
\begin{equation}
\frac{DS^{ij}}{ds}=P^i u^j-P^j u^i.
\label{motionspin}
\end{equation}

Putting (\ref{franikoenergymomentum}) and (\ref{franikospin}) into the conservation law (\ref{energymomentumconservation}) for tetrad energy--momentum density with (\ref{torsionvector}) gives
\[
(\mathfrak{e}\bar{\psi}\psi P_i u^j)_{;j}=2S^j_{\phantom{j}ki}(\mathfrak{e}\bar{\psi}\psi P_j u^k)+\frac{1}{2}R^{kl}_{\phantom{kl}ji}(\mathfrak{e}\bar{\psi}\psi S_{kl}u^j).
\]
Using $u^k\nabla_k=D/ds$, the four-velocity (\ref{velocity}), and the conservation of current (\ref{current}), this relation can be written as
\begin{eqnarray}
& & 2S^j_{\phantom{j}ki}(\mathfrak{e}\bar{\psi}\psi P_j u^k)+\frac{1}{2}R^{kl}_{\phantom{kl}ji}(\mathfrak{e}\bar{\psi}\psi S_{kl}u^j)=\mathfrak{e}\bar{\psi}\psi u^j \nabla_j P_{i}+P_i\nabla_j(\mathfrak{e}\bar{\psi}\psi u^j)=\mathfrak{e}\bar{\psi}\psi\frac{DP_i}{ds}+P_i\nabla_j(\mathfrak{e}\bar{\psi}\gamma^j\psi) \nonumber \\
& & =\mathfrak{e}\bar{\psi}\psi\frac{DP_i}{ds},
\nonumber
\end{eqnarray}
which is equivalent to
\begin{equation}
\frac{DP^i}{ds}=2S_{jk}^{\phantom{jk}i}P^j u^k+\frac{1}{2}R_{jkl}^{\phantom{jkl}i}S^{jk}u^l.
\label{motionmomentum}
\end{equation}
Equations (\ref{motionspin}) and (\ref{motionmomentum}) are the Mathisson--Papapetrou equations of classical motion of a particle with spin in a spacetime with curvature and torsion \cite{MP}.
Consequently, if a spinor satisfies the Dirac equation, then the corresponding particle with four-momentum $P_i$ given by an eigenvalue of $\hat{P}_i$ (\ref{momentum}), spin four-tensor $S_{ik}$ given by an eigenvalue of $\hat{S}_{ik}$ (\ref{spin}), and four-velocity $u^i$ (\ref{velocity}) satisfies the Mathisson--Papapetrou equations of motion.
These equations are supplemented by the orthogonality relation (\ref{orthogonality}).

In the absence of curvature and torsion, the equation of motion for the four-momentum (\ref{motionmomentum}) gives
$dP^i/ds=0$, so $P^i$ is constant.
Accordingly, integrating the equation of motion for the four-spin (\ref{motionspin}) gives the constancy of the total (orbital and spin) angular momentum four-tensor $x^i P^j-x^j P^i+S^{ij}=M^{ij}+S^{ij}$.\\

\noindent
{\bf Normalization}.\\
If the spinor mass is constant, then the square of the four-velocity (\ref{velocity}) of the spinor is constant.
To demonstrate that, we use the Mathisson--Papapetrou equations, which follow from (\ref{velocity}) applied to the conservation laws for the Dirac Lagrangian density.
Contracting (\ref{motionspin}) with $u_j$ and using (\ref{mass}) gives
\begin{equation}
P^i u^j u_j-mcu^i=\frac{DS^{ij}}{ds}u_j.
\label{proportionality}
\end{equation}
Differentiating (\ref{proportionality}) with $u_i D/ds$ and using $u_i DP^i/ds=0$, which follows from contracting (\ref{motionmomentum}) with $u_j$, gives
\[
P^i u_i\frac{D(u^j u_j)}{ds}-mcu_i\frac{Du^i}{ds}=u_i\frac{DS^{ij}}{ds}\frac{Du_j}{ds}.
\]
This relation is equivalent, using (\ref{mass}) and (\ref{orthogonality}), to
\[
\frac{1}{2}mc\frac{D(u^j u_j)}{ds}=-S^{ij}\frac{Du_i}{ds}\frac{Du_j}{ds}=0,
\]
and therefore
\[
u^j u_j=\alpha^2=\mbox{const}.
\]

The four-velocity (\ref{velocity}), which is $u^i=dx^i/ds$, can be normalized by scaling the affine parameter $s\to\tilde{s}=\alpha s$, so that $\tilde{u}^i=dx^i/d\tilde{s}=u^i/\alpha$ satisfies \cite{Wong}
\[
\tilde{u}^j\tilde{u}_j=1.
\]
The normalized four-velocity is orthogonal to the spin four-tensor: $S_{ik}\tilde{u}^k=0$.
Scaling the mass $m\to\tilde{m}=m/\alpha$ gives $\tilde{m}c=\tilde{u}^i P_i$, as in (\ref{mass}).
The relation (\ref{proportionality}) simplifies to
\begin{equation}
P^i-\tilde{m}c\tilde{u}^i=\frac{DS^{ij}}{d\tilde{s}}\tilde{u}_j,
\label{simplerproportionality}
\end{equation}
which gives the four-momentum of the spinor.
The Mathisson--Papapetrou equations are satisfied for the scaled quantities $\tilde{s}$ and $\tilde{u}^i$.\\

\noindent
{\bf Antisymmetry of spin tensor}.\\
The spin density (\ref{spindensity}) is equivalent to
\[
\mathfrak{S}^{ijk}=\frac{1}{2}i\hbar\mathfrak{e}\bar{\psi}\gamma^{[i}\gamma^j\gamma^{k]}\psi=\mathfrak{S}^{[ijk]},
\]
which is completely antisymmetric \cite{EC,Niko}.
Consequently, (\ref{franikospin}) gives
\[
S_{jk}u_i=-S_{ji}u_k.
\]
Contracting this relation with $u^i$ and using (\ref{orthogonality}) yields
\[
S_{jk}\alpha^2=-S_{ji}u^i u_k=0,\quad S_{ij}=0.
\]
This result does not mean that the spin density $\mathfrak{S}_{\mu\nu}^{\phantom{\mu\nu}i}$ vanishes, which would follow from (\ref{franikospin}).
It means that the pole approximation of a particle, which gives (\ref{franikospin}), must be extended to the pole-dipole description \cite{NSH} in order to account for the angular momentum in the equations of motion.
In that description, the four-tensor $S_{ik}$ appearing in the Mathisson--Papapetrou equations of motion (\ref{motionspin}) and (\ref{motionmomentum}) is replaced by the orbital angular momentum four-tensor $M_{ik}$ and spin-1/2 particles are spatially extended \cite{non}.\\

\noindent
{\bf Geodesic equation}.\\
In the pole description of a particle associated with a spinor wave function, the equations of motion (\ref{motionspin}) and (\ref{motionmomentum}) reduce to
\[
P^i u^j-P^j u^i=0,\quad \frac{DP^i}{ds}-2S_{jk}^{\phantom{jk}i}P^j u^k=0,
\]
where the four-momentum (\ref{simplerproportionality}) is proportional to the four-velocity:
\begin{equation}
P^i=mcu^i.
\label{momentumvelocity}
\end{equation}
Consequently, the equation of motion for the four-momentum becomes an equation for the four-velocity:
\begin{eqnarray}
& & \frac{Du^i}{ds}-2S_{jk}^{\phantom{jk}i}u^j u^k=\frac{du^i}{ds}+\Gamma^{i}_{jk}u^j u^k-2S_{jk}^{\phantom{jk}i}u^j u^k \nonumber \\
& & =\frac{du^i}{ds}+\mathring{\Gamma}^{i}_{jk}u^j u^k+2S^i_{\phantom{i}jk}u^j u^k=\frac{du^i}{ds}+\mathring{\Gamma}^{i}_{jk}u^j u^k \nonumber \\
& & =\frac{\mathring{D}u^i}{ds}=0, \nonumber
\end{eqnarray}
which is the geodesic equation \cite{LL2}.
A spinor particle in the pole approximation thus moves along a geodesic in a gravitational field.
The geodesic equation coincides with the autoparallel equation $Du^i/ds=0$ because the torsion tensor is completely antisymmetric, which follows from the Cartan equations (\ref{Cartan}) and the complete antisymmetry of the spin density.

Multiplying the first equation in (\ref{Dirac}) by $\bar{\psi}\gamma^\nu$ from the left and the second equation in (\ref{Dirac}) by $\gamma^\nu\psi$ from the right, and adding the two multiplied equations gives
\[
i\hbar e_\mu^i(\bar{\psi} \gamma^\nu\gamma^\mu \nabla_i\psi - \nabla_i\bar{\psi} \gamma^\mu\gamma^\nu \psi) = 2mc \bar{\psi}\gamma^\nu\psi.
\]
If $\psi$ represents a wave function with four-momentum $P_i$ and $\bar{\psi}$ represents a wave function with four-momentum $P'_i$, then using (\ref{momentum}) and $\gamma^\mu\gamma^\nu=\eta^{\mu\nu}+2G^{\mu\nu}$ yields
\begin{equation}
\bar{\psi}\gamma^\nu\psi = \frac{1}{2mc}
\bar{\psi}\psi(P^\nu+P'^\nu) + \frac{1}{mc}\bar{\psi}G^{\nu\mu}\psi(P_\mu- P'_\mu),
\label{Gordon}
\end{equation}
which is the Gordon decomposition of a spinor four-current $j^\nu=\bar{\psi}\gamma^\nu\psi$ \cite{QFT}.
If $\psi$ and $\bar{\psi}$ represent the same spinor, then $P'^\mu=P^\mu$, leading to
$\bar{\psi}\gamma^\nu\psi = \bar{\psi}\psi P^\nu/mc$.
Applying (\ref{velocity}) to this relation gives $P^\nu=mcu^\nu$ (\ref{momentumvelocity}), justifying the geodesic equation of motion.\\ \\ \\

{\bf 4. Equations for operators}\\ \\
Another justification for the formula (\ref{velocity}) for the four-velocity of a spinor comes from a consideration of the dynamics of a particle in the Heisenberg picture.
In nonrelativistic quantum mechanics, an operator $\hat{O}$ changes in time according to the Heisenberg equation of motion: $i\hbar(d\hat{O}/dt)=[\hat{O},\hat{H}]$, where $\hat{H}$ is the Hamiltonian of the system \cite{Dir}.
Wong \cite{Wong} showed that the covariant equations of motion in general relativity for a massive spin-1/2 particle can be obtained from a generalized Heisenberg picture, in which time is an operator and the dynamics of operators is given in terms of the proper time $\tau=s/c$ by $i\hbar(d\hat{O}/d\tau)=[\hat{O},\hat{H}]$, where the generalized, covariant Hamiltonian is
\[
\hat{H}=-\gamma^k\hat{P}_k c+mc^2 I_4.
\]
This dynamics is equivalent to the Schr\"{o}dinger picture: $i\hbar(\partial\psi/\partial\tau)=\hat{H}\psi$.

We show that the covariant equations of motion for the operators of the momentum four-vector and the spin four-tensor of a particle in the Einstein--Cartan gravity can be derived from the Heisenberg equation with the covariant derivative:
\begin{equation}
i\hbar\frac{D\hat{O}}{d\tau}=[\hat{O},\hat{H}].
\label{Heisenberg}
\end{equation}
The relation (\ref{covariant1}) for the covariant derivative operator gives, using (\ref{momentum}) and (\ref{spin}), a relation for the covariant four-momentum operator:
\[
\hat{P}_k=\hat{\pi}_k+\frac{1}{2}\omega^{\mu\nu}_{\phantom{\mu\nu}k}\hat{S}_{\mu\nu},
\]
where $\hat{\pi}_k=i\hbar\partial_k$ is the flat-space four-momentum operator.
Putting $\hat{O}=\hat{x}^i$ in (\ref{Heisenberg}) with the ordinary derivative over $\tau$ (because $\hat{x}^i$ is not a four-vector) and using $[\hat{x}^i,\hat{\pi}_k]=-i\hbar\delta^i_k$, where $\delta^i_k$ is the Kronecker tensor, gives
\[
\hat{u}^i=\frac{d\hat{x}^i}{ds}=\gamma^i.
\]
The expectation value of this relation is $u^i=\bar{\psi}\hat{u}^i\psi/\bar{\psi}\psi=\bar{\psi}\gamma^i\psi/\bar{\psi}\psi$, agreeing with the formula (\ref{velocity}).

The relation $\nabla_i\gamma^k=0$ gives $[\nabla_i,\gamma^k]=[\hat{P}_i,\gamma^k]=0$ and also $[\hat{P}_j,G_{ik}]=0$.
For $\hat{O}=\hat{S}^{ij}$ in (\ref{Heisenberg}), we get $D\hat{S}_{ik}/d\tau=-[G_{ik},\gamma^j]\hat{P}_j c-\gamma^j[G_{ik},\hat{P}_j]c$.
Using $[\gamma^j,G_{ik}]=\delta^j_i\gamma_k-\delta^j_k\gamma_i$ and $\gamma^i=\hat{u}^i$ therefore yields
\[
\frac{D\hat{S}_{ik}}{ds}=\hat{u}_k\hat{P}_i-\hat{u}_i\hat{P}_k,
\]
whose expectation value is equivalent to the Mathisson--Papapetrou equation (\ref{motionspin}).
For $\hat{O}=\hat{P}^i$ in (\ref{Heisenberg}), we get $i\hbar(D\hat{P}_i/d\tau)=-[\hat{P}_i,\gamma^k]\hat{P}_k c-\gamma^j[\hat{P}_i,\hat{P}_j]c$.
The commutator of the four-momentum operators follows from the commutator of the covariant derivatives of a spinor and the formula for the curvature spinor:\footnote{
In the presence of torsion, the four-momentum operators do not commute.
Consequently, the integration over four-momentum in loop Feynman diagrams must be replaced with the summation over four-momentum eigenvalues \cite{torsional}.
Divergent integrals are changed into convergent sums, eliminating ultraviolet divergence in quantum field theory and making renormalization finite.}\\
\[
[\hat{P}_i,\hat{P}_j]=\frac{1}{2}i\hbar R^{kl}_{\phantom{kl}ij}\hat{S}_{kl}+2i\hbar S^k_{\phantom{k}ij}\hat{P}_k.
\]
Using $\gamma^i=\hat{u}^i$ therefore yields
\[
\frac{D\hat{P}_i}{ds}=-\frac{1}{2}\hat{u}^j R^{kl}_{\phantom{kl}ij}\hat{S}_{kl}-2\hat{u}^j S^k_{\phantom{k}ij}\hat{P}_k,
\]
whose expectation value is equivalent to the Mathisson--Papapetrou equation (\ref{motionmomentum}).\\ \\

{\bf 5. Pilot-wave quantum mechanics}\\ \\
\noindent
{\bf de Broglie--Bohm quantum mechanics}.\\
In the nonrelativistic limit, $v\ll c$, the Dirac equation becomes the Schr\"{o}dinger equation:
\[
i\hbar\frac{\partial\psi}{\partial t}=-\frac{\hbar^2}{2m}\nabla^2\psi+U\psi,
\]
where $U$ is the potential energy.
The wave function $\psi$ is a complex number, which can be written as $\psi=R\exp(iS/\hbar)$, where $R$ and $S$ are real functions of the coordinates.
The imaginary part of the Schr\"{o}dinger equation gives $\partial R/\partial t=-(1/2m)(R\nabla^2 S+2{\bm\nabla}R\cdot{\bm\nabla}S)$.
Using here the Born rule for the probability density, $\rho=|\psi|^2=R^2$, gives the equation of continuity, $\partial\rho/\partial t+{\bm\nabla}\cdot(\rho{\bf v})=0$,
for the velocity field:
\begin{equation}
{\bf v}=\frac{1}{m}{\bm\nabla}S.
\label{guiding}
\end{equation}
Because ${\bf p}=m{\bf v}$, the function $S$ coincides with the action for the particle, for which ${\bf p}={\bm\nabla}S$.

The real part of the Schr\"{o}dinger equation gives
\[
\frac{\partial S}{\partial t}=-\frac{1}{2m}({\bm\nabla}S)^2+U+Q,
\]
which is the Hamilton--Jacobi equation for the action $S$ with an additional energy term called the quantum potential \cite{Bohm}:
\[
Q=-\frac{\hbar^2}{2m}\frac{\nabla^2 R}{R}.
\]
The velocity field (\ref{guiding}), which depends on $Q$ through $S$, can be integrated over time to calculate localized trajectories of particles.
Those trajectories can explain interference patterns in the double-slit experiment and scattering from square barriers and square wells \cite{exp}; the apparent random distribution of the final positions of particles comes from the lack of knowledge about their exact initial positions.
Therefore, the pilot-wave theory is a deterministic interpretation of quantum mechanics, which gives the same predictions as the standard, probabilistic Copenhagen interpretation, avoiding wave function collapse and the measurement problem \cite{BH}.\\

\noindent
{\bf Relativistic pilot-wave theory}.\\
In the nonrelativistic limit, the Gordon decomposition of a spinor four-current (\ref{Gordon}) with the relation (\ref{velocity}) give the velocity field (\ref{guiding}).
In this limit, the formula (\ref{velocity}) thus becomes the guiding equation for the velocity of a particle in the de Broglie--Bohm pilot-wave interpretation of quantum mechanics.
Consequently, we propose that the four-velocity (\ref{velocity}) provides a general-relativistic generalization of the guiding equation of the pilot-wave quantum mechanics.
It extends to a curved spacetime the equations of relativistic pilot-wave theory in a flat spacetime, which have been derived using Clifford algebras \cite{HC} or a privileged foliation of spacetime \cite{rel}.

The wave function determines the four-velocity field for particles.
The wave function of the matter tells the spacetime how to curve according to the Einstein and Cartan field equations for the densities (\ref{spindensity}) and (\ref{energymomentumdensity}), and the spacetime governs the propagation of the wave function.
The wave function also pilots the motion of the matter according to (\ref{velocity}).
Therefore, the spacetime tells the matter how to move using pilot waves, possibly unifying these entities.

Equation (\ref{velocity}) can be written as $\bar{\psi}u^i\psi=\bar{\psi}\gamma^i\psi$, which shows that the bilinear vectors constructed from $u^i$ and $\gamma^i$ are equal.
It interesting to note that replacing $\gamma^i$ with $u^i$ in the Dirac equation (\ref{Dirac}) gives
\[
\frac{D\psi}{ds}=u^i \nabla_i \psi = (mc/i\hbar)\psi,
\]
where $D$ denotes a covariant differential.
This relation integrates to
$\psi\sim \exp(-imc\int ds/\hbar)$.
Comparing this relation with a semiclassical relation $\psi\sim \exp(iS/\hbar)$ gives the action for a free particle: $S=-mc\int ds$ \cite{LL2}.\\

\noindent
{\bf Wave packets and nonlinearity}.\\
In the nonrelativistic pilot-wave quantum mechanics, the motion of particles guided by Gaussian wave packets is accelerated \cite{Pan}.
The same result is expected for the relativistic theory with the four-velocity (\ref{velocity}).
In the Einstein--Cartan theory, the covariant derivative in the Dirac equation (\ref{Dirac}) is linear in torsion and the torsion tensor for a Dirac field is quadratic in spinor fields because of the spin density (\ref{spindensity}) and the Cartan equations (\ref{Cartan}).
Consequently, the Dirac equation is cubic (nonlinear) in spinor fields \cite{Niko,EC}:
\[
i\hbar \gamma^\mu e_\mu^i\mathring{\nabla}_i\psi + \frac{3}{8}\kappa(\hbar c)^2(\bar{\psi}\gamma_\mu\gamma^5\psi)\gamma^\mu\gamma^5\psi = mc\psi,
\]
where $\gamma^5=i\gamma^0\gamma^1\gamma^2\gamma^3=\left( \begin{array}{cc}
    0 & I_2 \\
    I_2 & 0 \end{array} \right)$.
Therefore, the superposition principle in quantum mechanics is violated by torsion and wave packets do not satisfy the Dirac equation like a wave with a determinate value of the four-momentum.
The nonlinearity of the Dirac equation in the Einstein--Cartan theory also removes the indeterminacy of a wave function, which in general relativity can be multiplied by an arbitrary constant (normalized) because the Dirac equation without torsion is linear.\\

{\bf 6. Summary}\\ \\
In this work, we proposed that the relativistic wave--particle duality relation (\ref{velocity}) provides a general-relativistic generalization of the guiding equation in the pilot-wave interpretation of quantum mechanics.
We showed the validity of this relation for a free particle related to a plane wave in flat spacetime. 
We used this relation and the covariant conservation laws for the spin and energy--momentum tensors to show that if a spinor wave function satisfies the Dirac equation in spacetime with curvature and torsion, then the corresponding particle satisfies the Mathisson--Papapetrou equations, which reduce to the geodesic equation of motion.

We also showed that the operator form of these equations and the duality relation can be derived from the Heisenberg equation of motion with the covariant derivative with respect to the proper time, extending the work of Wong \cite{Wong} to spacetime with torsion.
All these novel results give a further validation of the equivalence between the classical and quantum dynamics of massive spin-1/2 particles in the gravitational field.\\

{\bf Acknowledgments}\\ \\
I am grateful to my Parents, Bo\.{z}enna Pop{\l}awska and Janusz Pop{\l}awski, for supporting this work - N P.

\end{document}